\documentclass[]{spie}  

\usepackage{amsmath,amsfonts,amssymb}
\usepackage{graphicx}
\usepackage[colorlinks=true, allcolors=blue]{hyperref}

\title{Astrophotonic Solutions for Spectral Cross-Correlation Techniques}

\author[a,b]{Suresh Sivanandam}
\author[c]{Ross Cheriton}
\author[d,a]{Polina Zavyalova}
\author[d]{Peter R. Herman}
\author[a,b]{Emily Deibert}
\author[c,e]{Erin Tonita}
\author[c,f]{Volodymyr Artyshchuk}
\author[g]{Ernst de Mooij}
\author[c]{Siegfried Janz}
\author[h]{Adam Densmore}
\affil[a]{David A. Dunlap Department of Astronomy and Astrophysics, University of Toronto, Toronto, Canada}
\affil[b]{Dunlap Institute for Astronomy and Astrophysics, University of Toronto, Toronto, Canada}
\affil[c]{Advanced Electronics and Photonics, National Research Council of Canada, Ottawa, Canada}
\affil[d]{Edward S. Rogers Sr. Department of Electrical \& Computer Engineering, University of Toronto, Toronto, Canada}
\affil[e]{University of Ottawa, Ottawa, Canada}
\affil[f]{Carleton University, Ottawa, Canada}
\affil[g]{Astrophysics Research Centre, School of Mathematics and Physics, Queen's University Belfast, Belfast, United Kingdom}
\affil[h]{Herzberg Astronomy and Astrophysics, National Research Council of Canada, Victoria, Canada}

\authorinfo{Further author information: (Send correspondence to S.S.)\\S.S.: E-mail: suresh.sivanandam@utoronto.ca, Telephone: 1 416 978 6550}

\begin{document} 
\maketitle

\begin{abstract}
Using photonic devices, we developed a new approach to traditional spectroscopy where the spectral cross-correlation with a template spectrum can be done entirely on-device. By creating photonic devices with a carefully designed, modulated transmission spectrum, the cross-correlation can be carried out optically without requiring any dispersion, vastly simplifying the instrument and reducing its cost. The measured correlation lag can be used for  detecting atomic/molecular species within and determining the radial velocity of a particular astrophysical object. 

We present an overview of two design approaches that are currently being developed that use different photonic platforms: silicon and fibre-based photonics. The silicon photonic approach utilizes ring resonators that can be thermo-optically modulated to carry out the cross-correlation. The fibre approach uses customized fibre Bragg gratings (FBGs) with transmission spectra that can be strain-modulated. Both approaches have been able to detect molecular gas in a lab setting, and we are now in the process of on-sky testing. 

Lastly, we discuss the future for these types of devices as their simplicity opens up the possibility of developing low-cost, purpose-built multi-object or integral field spectroscopic instruments that could make significant contributions to scientific programs requiring stellar RV measurements and exoplanet detections.
\end{abstract}


\section{Introduction}
\label{sec:intro}  

Spectroscopic measurements are fundamental for understanding the properties of astronomical objects. These measurements are essential for determining these objects' kinematics, composition, and excitation states. Spectral cross-correlation techniques are often used to measure precise radial velocities and determine composition of stars and exoplanets. These techniques require spectra from high dispersion echelle spectrographs, which are complex and costly to design and construct. Modern echelle spectrographs require sophisticated optical designs, large echelle gratings, and large format image sensors to reach their design goals of reaching high spectral resolutions of R $>20,000$ over a broad wavelength range. This is needed for both precise stellar radial velocity (RV) measurements and exoplanet atmospheric characterization\cite{2021A&A...645A..96P,2020MNRAS.498.5684D}. 

For these measurements, astrophysical information is extracted by cross-correlating spectra rich in atomic and/or molecular features with models to extract radial velocities (RVs) and compositions. In the case of stellar measurements, either binary masks or spectral models are constructed to match the physical conditions of the stellar photosphere to cross-correlate with. Similarly in exoplanet transit or emission spectroscopy, a multitude of exoplanet atmospheric models are cross-correlated with the stellar transit spectrum to determine the composition of its atmosphere. 

Attempts to reduce the cost and complexity of high dispersion spectrographs date back several decades primarily due to the lack of availability of large format image sensors\cite{1967ApJ...148..465G,1982PASP...94.1017F}. These instruments measured RVs of stellar objects by using a traditional high dispersion spectrograph, which used a specialized focal plane mask and a photodetector to measure the transmission through the mask, instead of a large format detector. The focal plane mask was designed to mimic the stellar spectral features of the object being studied. By translating the mask by known amounts and measuring the photometer output, the radial velocity of the object can be measured. This technique is in effect carrying out an optical cross-correlation between the stellar spectrum and specialized mask. Despite being several decades old, these instruments were able to reach RVs better than 1 km s$^{-1}$\cite{1982PASP...94.1017F}. Nonetheless, they share a similar level of optical complexity when compared to modern echelle spectrographs. Astrophotonics offers a new and more elegant solution for carrying out the optical cross-correlation at lower complexity and cost.  

The field of astrophotonics has been growing dramatically over the past decade. Astrophotonic devices offer new approaches to transitional astronomical instrumentation. These include miniature spectrometers\cite{2021OExpr..2924947S,2021SPIE11819E..0IG}, customizable notch filters\cite{2015SPIE.9507E..0CE,2021OExpr..2915867P}, multi-mode to single mode converters\cite{2010OExpr..18.8430L} (photonic lanterns). Due to the scale of these astrophotonic instruments, they are much more compact and cheaper than one-of-a-kind traditional instruments that use bulk optics. A recent review\cite{2021A&ARv..29....6M} presents the current state-of-the-art in this field. Astrophotonics now offers a modern solution to the decades old cross-correlation spectrograph concept without any need for dispersion. Through the design of customizable notch filters that can be modulated, we have developed a cross-correlation spectrograph, a correlation sensor in short, that is tailor made for specific scientific problems. The solution is entirely non-dispersive and does not require complex optics and can be fabricated on a multitude of platforms (e.g. silicon photonics, fibres).  

In Section \ref{sec:technique}, we discuss our proposed optical cross-correlation technique. In Section \ref{sec:implementation}, we present specifics of the implementation of this technique. In Section \ref{sec:future}, we highlight future prospects for this method, which could simplify highly multiplex spectroscopy. Finally, in Section \ref{sec:summary}, we summarize our instrument concept. 

\section{Optical Cross-Correlation Technique}
\label{sec:technique}
To develop the technique, the process begins with the identification of the scientific problem that needs to be solved, which requires spectroscopy. Next, the photonic device will need to be customized for that program by first developing the required spectral template needed for the cross-correlation. As an example, we focus on a frequent scientific need to measure stellar radial velocities. Common science cases include the measurement of kinematics of resolved stars within our galaxy and nearby dwarf galaxies to constrain their mass distributions. For this program, we target the Calcium Triplet, which is a particularly prominent feature in bright giant stars that are often used as targets. Located at around 0.85 $\mu$m, the calcium triplet (CaT) forms very strong absorption lines that is frequently used for RV measurements. To illustrate our technique, we have simulated a custom three notch transmission filter that matches the CaT lines (Figure \ref{fig:correlation} Left Panel). In our concept, we then modulate this notch filter across the CaT lines and measure the transmitted intensity (Figure \ref{fig:correlation} Right Panel). The transmitted intensity is the cross-correlation between the notch filter and the input spectrum:
\begin{equation}
    C(v) =  S(\lambda) \star T(\lambda),
\end{equation}
where $v$ = $\Delta\lambda/\lambda_0$*c, $S$ is the spectrum of the object, $T$ is the transmission profile of the notch filter, $C$ is the observed cross-correlation lag, $\Delta\lambda$ is the wavelength offset from the rest wavelength when cross-correlating, and $\lambda_0$ is where maximum correlation is achieved with an object at rest. $\lambda_0$ corresponds to specific spectral feature you are correlating with. There is a trade-off between the width of each notch of the filter and the time to scan across the features in order to obtain sufficient signal-to-noise ratio (SNR) in the correlation measurement. This needs to be chosen carefully for a particular scientific problem.

\subsection{Measuring Radial Velocity}
\begin{figure} [ht]
   \begin{center}
   \begin{tabular}{cc} 
   \includegraphics[height=5.5cm]{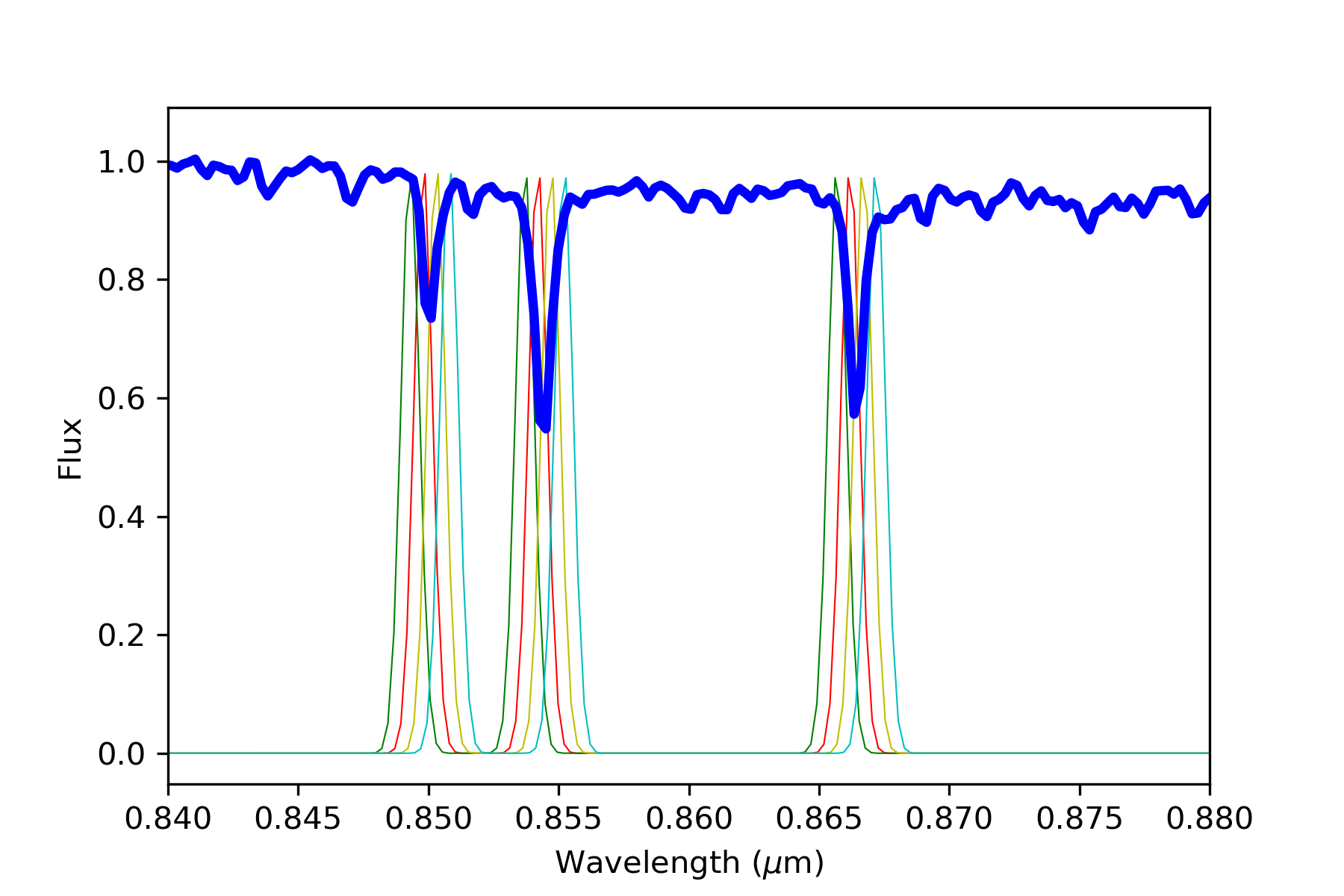} & \includegraphics[height=5.5cm]{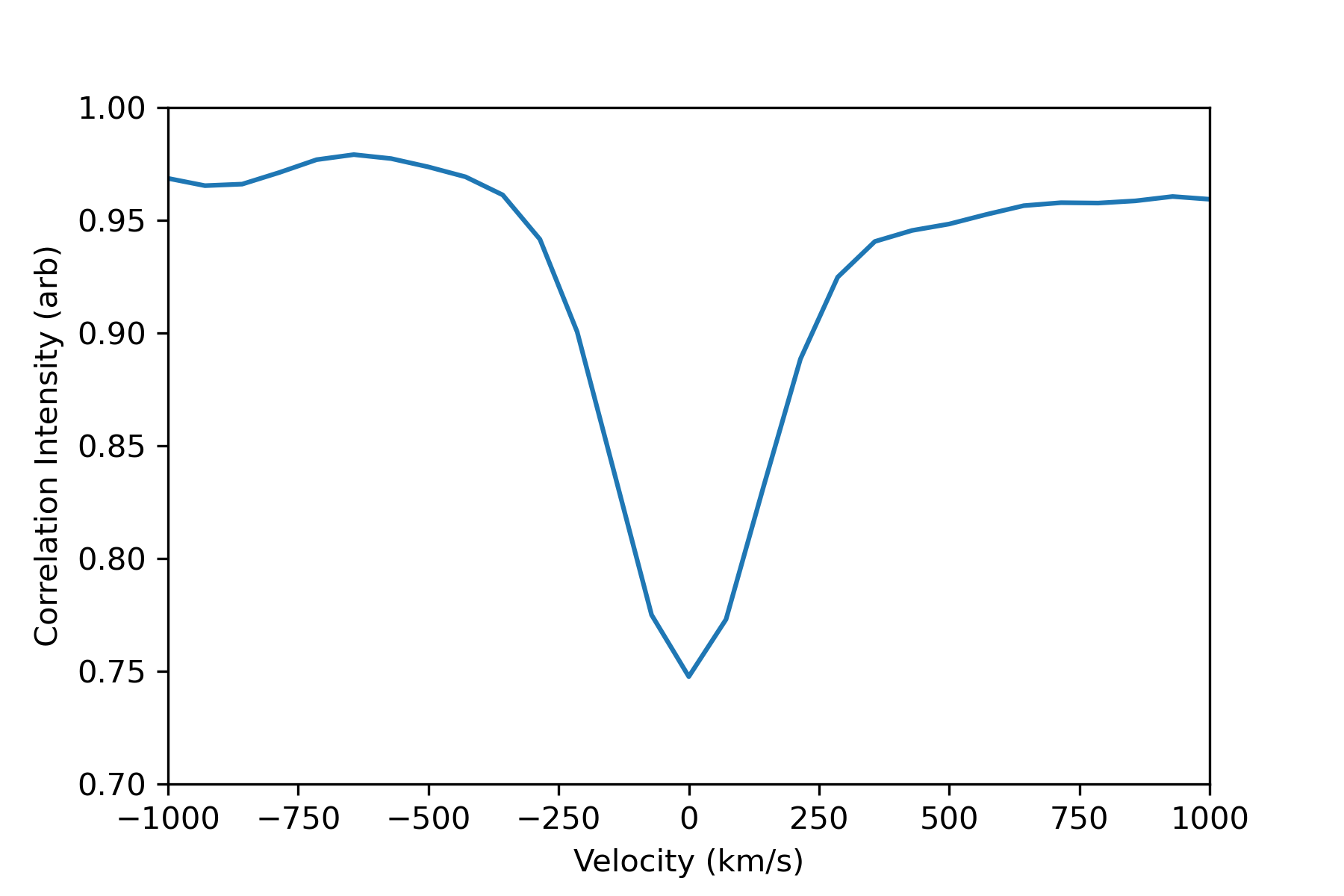}
   \end{tabular}
   \end{center}
   \caption[example] 
   { \label{fig:correlation} 
{\it Left:} The blue curve is a spectrum of a G4III giant obtained from the NASA IRTF SPEX spectral library. The Calcium Triplet (CaT) lines are prominent. The thinner curves are the filter notches designed to correlate with the CaT features. These notch filters are modulated across the CaT features, and the cross-correlation result is shown in the right panel. {\it Right:} Simulated intensity output from correlation sensor scaled by radial velocity offset of modulation. The dip in intensity indicates where the correlation is the strongest. This position can be used to measure the objects radial velocity and/or the existence of that spectral feature. The SPEX spectrum has been shifted to have zero radial velocity, which is indicated by the correlation signal close to no radial velocity. }
   \end{figure}

In this illustrative case, this spectrum of the object, G4III giant star, was taken from the NASA IRTF SPEX spectral library\cite{2009ApJS..185..289R}.  As shown in Figure \ref{fig:correlation}, when the notch filter aligns with the absorption features, there is a significant dip in intensity through the filter, indicating that we have reached the appropriate RV for the object. The width of the notch filter was chosen to be similar to the width of the individual CaT absorption lines in the spectrum. In this case, the spectral features are limited by the IRTF SPEX spectrograph\cite{2003PASP..115..362R}, which has a spectral resolving power of $R\sim2,000-2,500$, requiring a FWHM of each notch filter to have a full-width-half-maximum (FHWM) of 0.35 nm. This FWHM is larger than the typical line width of CaT features ($0.15-0.2$ nm) in red giant stars. In general, the observations done by our cross-correlation instrument will only be ultimately limited by the intrinsic line width of the astrophysical source and therefore the filters need to be designed accordingly. By measuring the position of the dip in the intensity after modulation, the RV of the source can be determined.   

\subsection{Measuring Composition}
\begin{figure} [ht]
   \begin{center}
   \begin{tabular}{cc} 
   \includegraphics[height=13cm]{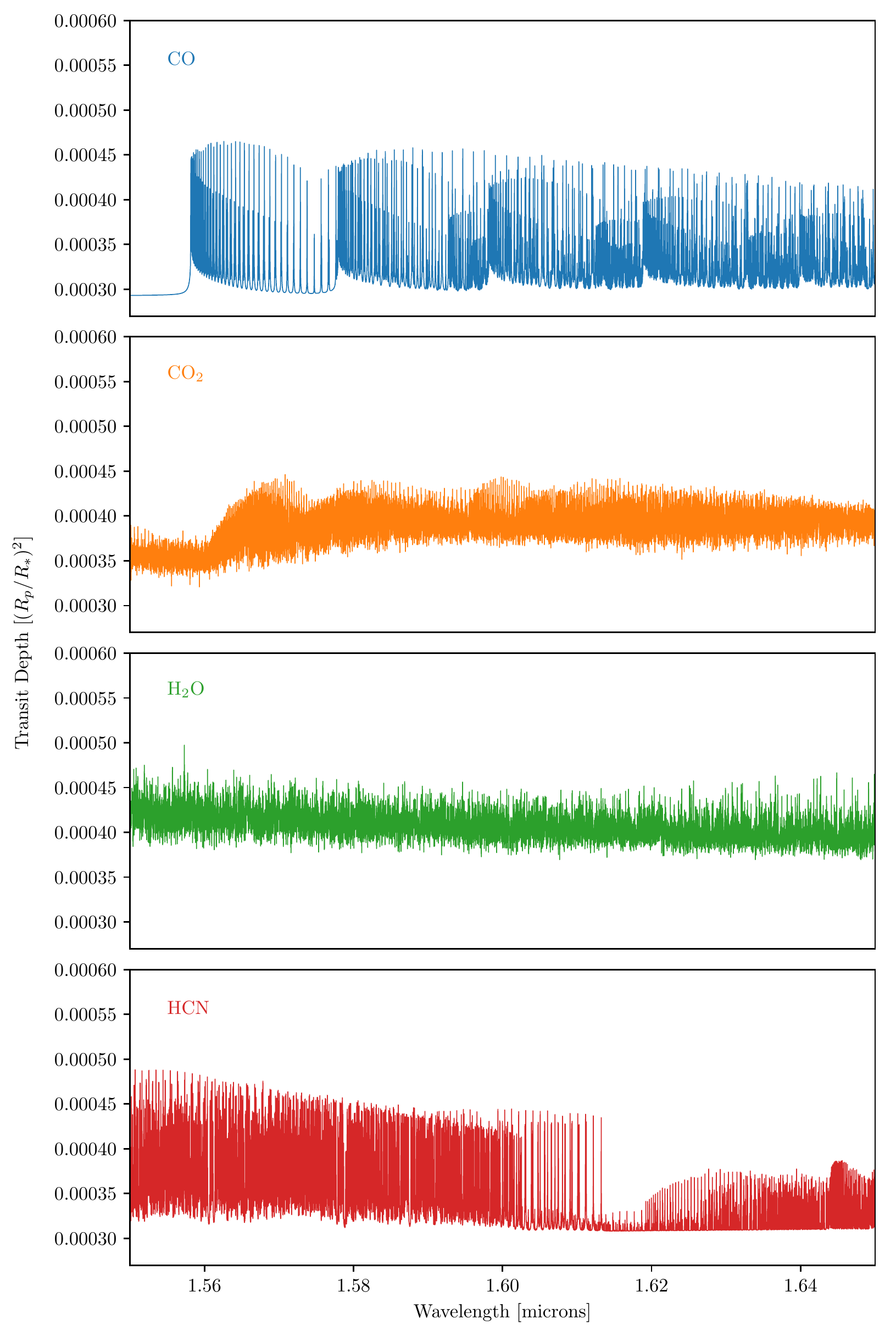} 
   \end{tabular}
   \end{center}
   \caption[example] 
   { \label{fig:exoplanet} 
Expected transit depth of a hot super-Earth like 55 Cancri e within a small region of $H$-band. The contributions to the transit depth from individual molecules are shown, which reveal a forest of absorption lines. A planet with no atsmophere will produce a transit depth of only 0.0003. The line distribution is a unique fingerprint for each molecule. A customized correlation sensor can be designed for each molecule in order to search for its existence within an exoplanet's atmosphere.}
   \end{figure}

 Another exciting direction is the measurement of exoplanet atmospheres through detection of molecular species in the optical and infrared. At this time, this is typically accomplished by cross-correlating model atmospheric templates with spectra obtained from high dispersion spectrographs\cite{2019AJ....157..114B}. Due to their lower temperatures, molecules are abundant in exoplanet atmospheres and consequently contain a forest of absorption lines that could be correlated with a specialized filter. This includes molecules such as O$_2$, H$_2$O, CH$_4$, CO, and CO$_2$. To illustrate this, some of the molecular features that can be observed in the infrared $H$-band for a transiting hot super-Earth, 55 Cancri e, are illustrated in Figure \ref{fig:exoplanet}. For reference, this planet has a temperature of 2000K with a radius of 1.95 R$_\oplus$ and its star is 0.98 R$_\odot$\cite{2018ApJ...860..122C}. The importance of a molecule is highly wavelength dependent and will need to be specifically targeted based on scientific program of interest. The correlation filter is particularly well-suited for molecules with pseudoperiodic features like CO and HCN in this case where there a few well-resolved lines that can be easily targeted. The principle of measurement is the same as for RVs. If the planet's orbital velocity as well as location in its orbit is known, the SNR of planet correlation signal can be increased by shifting and stacking the correlation signal to account for the planet's changing line-of-sight velocity. 

\section{Instrument Concept}
\label{sec:implementation}
As discussed, the instrument concept involves the construction of an astrophotonic correlation filter that contains customizable notch filters, which are generated through optical interference within a single mode device. These filters can be reliably and repeatably modulated, allowing our instrument to scan across the features of interest. The modulation technique is specific to the implementation, and we discuss them in more detail in the following subsections. While the modulation yields an effective loss of throughput compared to a dispersive spectrograph, the simple non-dispersive nature of these devices means that there is a significant throughput advantage over traditional echelle spectrographs (70\% for a photonic device versus 15\% for an echelle spectrograph). 

 \begin{figure} [ht]
   \begin{center}
   \begin{tabular}{cc} 
   \includegraphics[height=8cm]{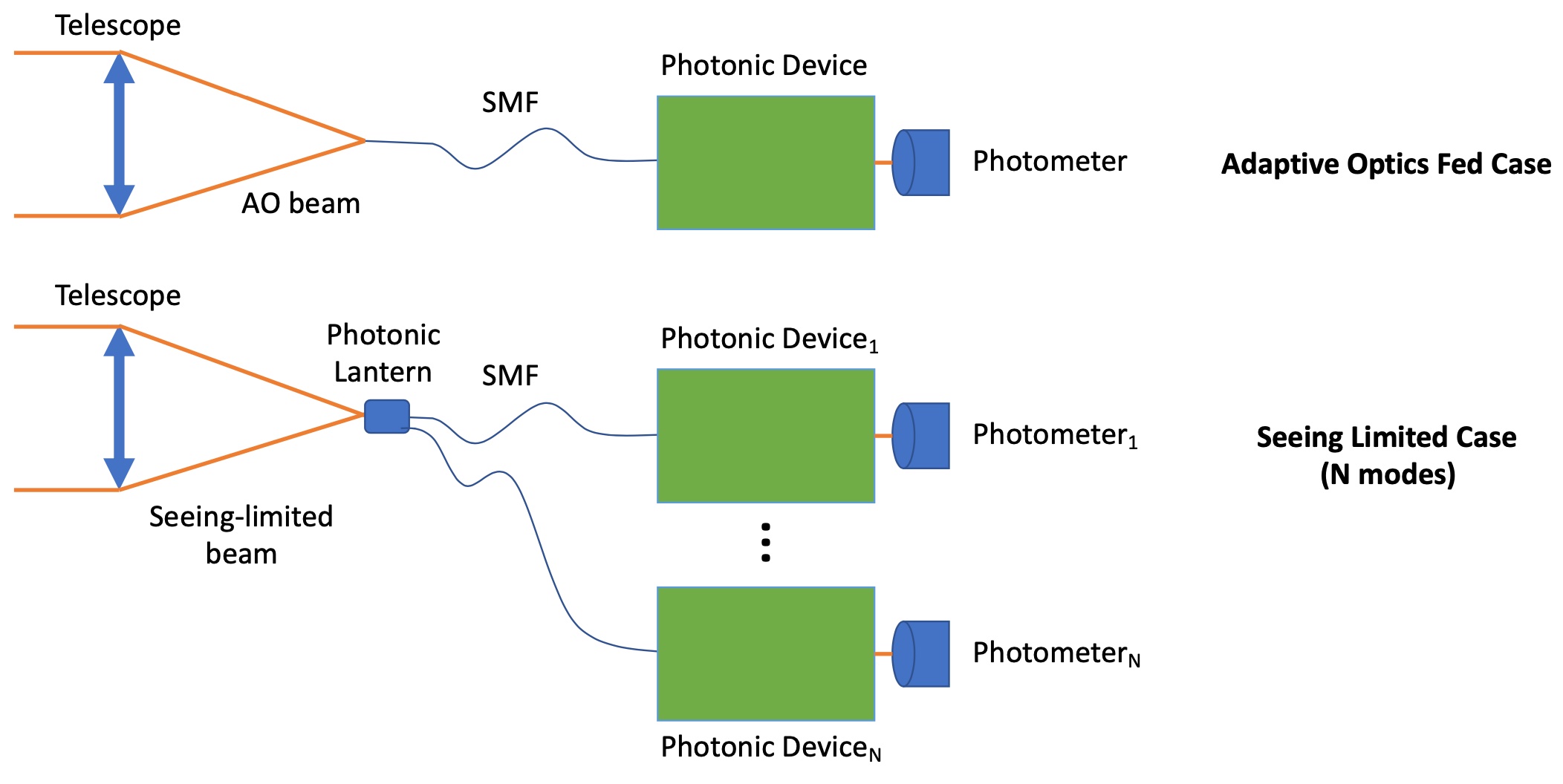}
   \end{tabular}
   \end{center}
   \caption[example] 
   { \label{fig:instrument} 
This figure shows the general instrument concept for two different cases: diffraction-limited and seeing-limited. The expectation is that the light from the object will be coupled from the telescope focal plane using an optical fibre, though that is not strictly necessary if we use integrated photonics. In the diffraction-limited case, a single mode fibre (SMF) couples the light into the photonic device. The output is then captured by a photometer, which could simply be a photodiode. In the seeing-limited case, light is fed into a multi-mode fibre which is split into multiple SMFs via a photonic lantern, which each feed individual photonic devices. N photonic devices and photometers are required for N modes.}
   \end{figure} 

The overall concept is presented in Figure \ref{fig:instrument}. Telescope light is fed via a single mode fibre (SMF) to a single device if it is diffraction-limited (e.g. with adaptive optics correction) or a photonic lantern that couples the beam into multiple SMFs, which each individually feed a device. The output of the device is either the integral of the either the reflected or transmitted spectrum of the correlation sensor, or both. The output is then coupled into a photometer, which is subsequently read out. If multiple devices are read simultaneously, they can feed multiple photometers or a linear sensor array. Thanks to the correlation being done optically, we only require a single photometer for each device, removing the need for large format sensors. This technique has inherent advantages over traditional methods especially for scientific programs that require a significant amount of spectroscopy, such as massively multiplexed spectroscopy, or for lightweight purpose-built instruments required for space applications. The overall cost and footprint is substantially lower and the devices can be easily replicated. 

To demonstrate this idea, we have developed two approaches that use either the silicon photonics or optical fibres to implement the correlation sensor. For an initial lab demonstration, we focused on the telecommunication band within the $1.5-1.6$ $\mu$m range. We chose a gas that is rich in pseudoperiodic lines and easy to work with in a lab environment: CO$_2$. The gas was also chosen on the basis that it can also be detected in the Earth's atmosphere and Venus in future experiments.

\subsection{Silicon Photonics}
\begin{figure} [ht]
   \begin{center}
   \begin{tabular}{cc} 
   \includegraphics[height=6cm]{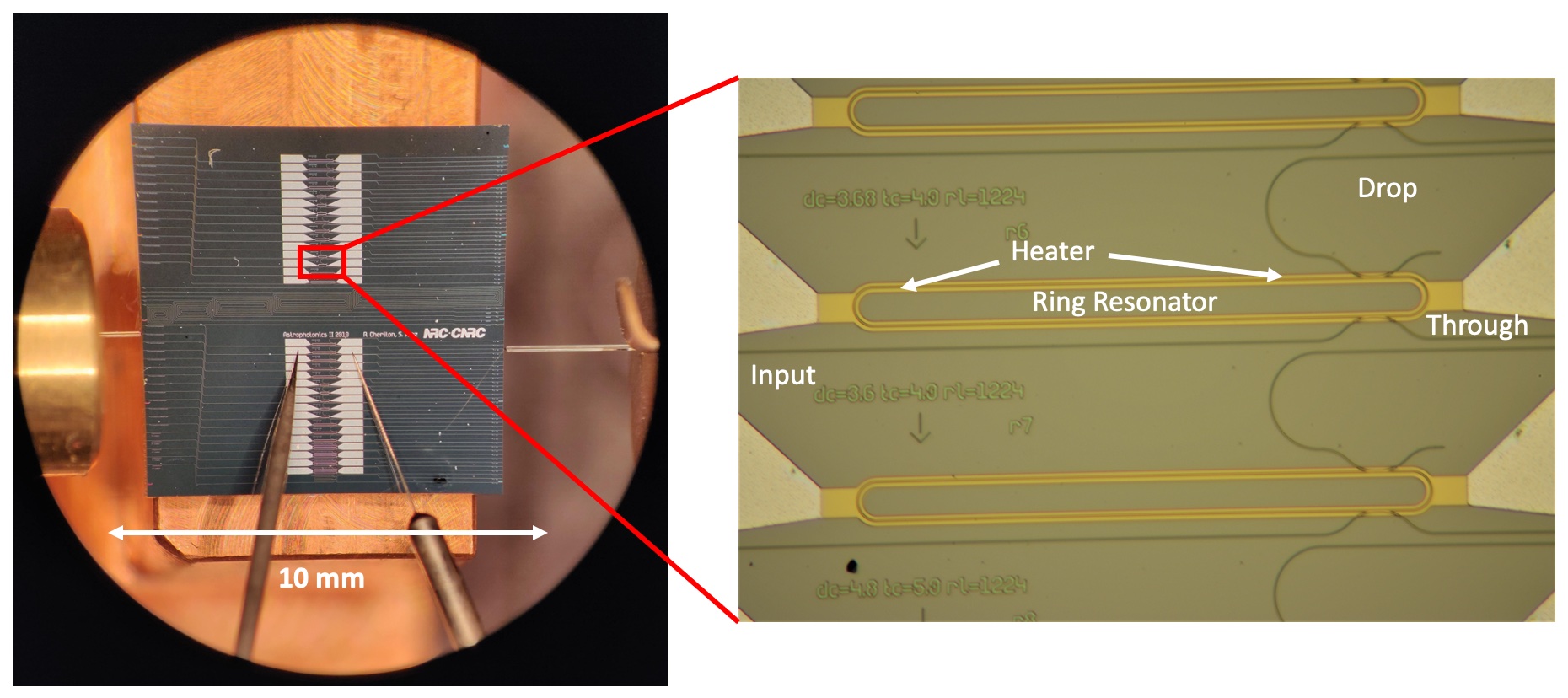}
   \end{tabular}
   \end{center}
   \caption[example] 
   { \label{fig:siphotonics} 
A microscope image of the silicon photonic chip with multiple test ring resonators. The device is only 8.8 mm on a side. Each ring resonator can be coupled separately with a tapered SMF. The magnified view of the ring resonator shows a view of the input, through (reflection), and drop (transmission) ports of the resonator as well as the the heating element deposited on top the resonator loop. There are electrical pads on each adjacent end of the heater to connect to a power supply that modulates the heater temperature. }
   \end{figure}

Discussed in further detail in our previous work\cite{2020OExpr..2827951C,2021ApOpt..6010252C}, our design uses the silicon-on-insulator photonics platform to implement the filter. With the intial batch of devices, we tested a number of gases, which included HCN and CO$_2$. Thanks to the pseudoperiodic nature of these gas absorption features, we used a ring resonator, which has equally spaced resonances, as our correlation filter. This approach was chosen for its simplicity in implementation. While the resonances do not fully match the CO$_2$ features, there is sufficient overlap of up to ten absorption lines at 1.58$\mu$m to obtain a strong correlation signal. We show the device in Figure \ref{fig:siphotonics}. The device is edge coupled by an SMF and consists of multiple ring resonators, which were designed for different gases. The modulation is achieved by a heater patterned on the ring resonator, which is also shown in the same figure. By using the thermooptic effect in silicon waveguide modes, we are able to modulate the ring resonator resonances across the CO$_2$ lines.

\begin{figure} [ht]
   \begin{center}
   \begin{tabular}{cc} 
   \includegraphics[height=7cm]{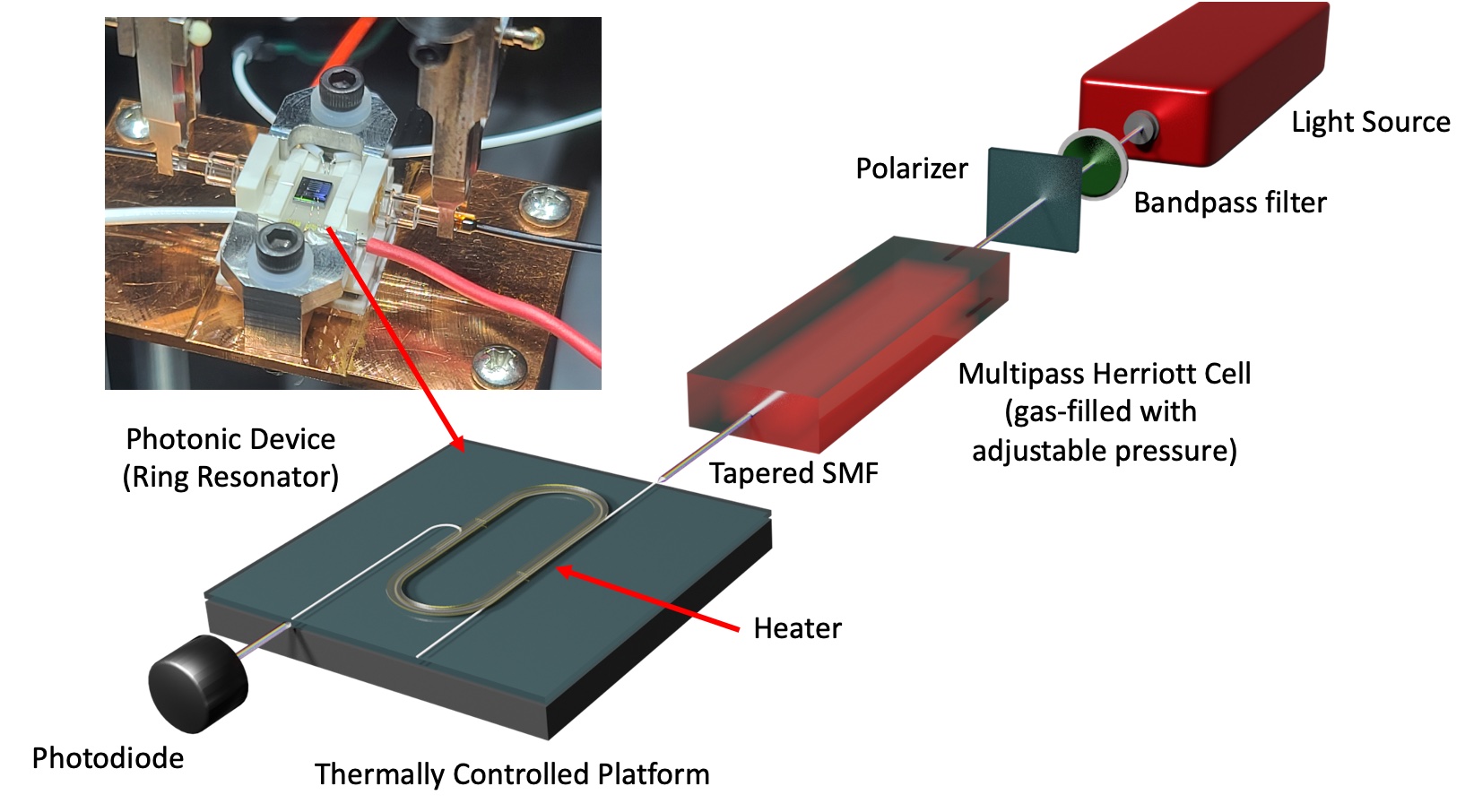}
   \end{tabular}
   \end{center}
   \caption[example] 
   { \label{fig:siexp} 
The lab experimental setup that was used to measure the photonic device's ability to detect and measure CO$_2$ is shown here. A multipass Herriott cell was used to test the sensitivity of the device to changes in CO$_2$ absorption by changing the pressure of the gas. The inset image shows the actual setup where the chip is set on a temperature controlled base. The chip is held slightly above ambient temperature to mitigate any impact from environmental temperature fluctuations. The power supply connections are wirebonded to the device.
}
   \end{figure}

The overall experimental setup is shown in Figure \ref{fig:siexp}. The photonic chip is placed on a temperature controlled stage to maintain a fixed base temperature while the current of ring resonator heaters is modulated to correlate with the gas being measured. A photodiode observes the output on the drop or through port of the resonator. As an additional method to reduce systematic errors, a lock-in amplifier is used to drive the heaters\cite{2021ApOpt..6010252C}. With this setup, we were able to measure variations in the CO$_2$ column density at the 10 parts-per-million level. The CO$_2$ column was adjusted by changing the pressure of the gas within the gas cell. As the next step, the input fibre was coupled to a fibre collimator and pointed at the Sun. We were able to successfully detect telluric CO$_2$ on-sky with this experiment.  
   
Our overall goal is to demonstrate the application on planetary and astrophysical sources. To this end, we have two experiments underway, one that couples light from a small astronomical telescope into the fibre (Tonita et al., this conference) and another that couples light from an adaptive optics system implemented on a 1.2-meter telescope. Currently, the overall throughput of our experimental system is low and our next steps are to improve the coupling efficiency. Additionally, we are also evaluating the silicon nitride (SiN) platform, which will allow us to develop devices that operate shortward of the silicon bandgap of 1.1 $\mu$m, allowing us to target other interesting molecules like O$_2$ in the visible regime.  
   
\subsection{Fibre Bragg Gratings}
\begin{figure} [ht]
   \begin{center}
   \begin{tabular}{cc} 
   \includegraphics[height=7cm]{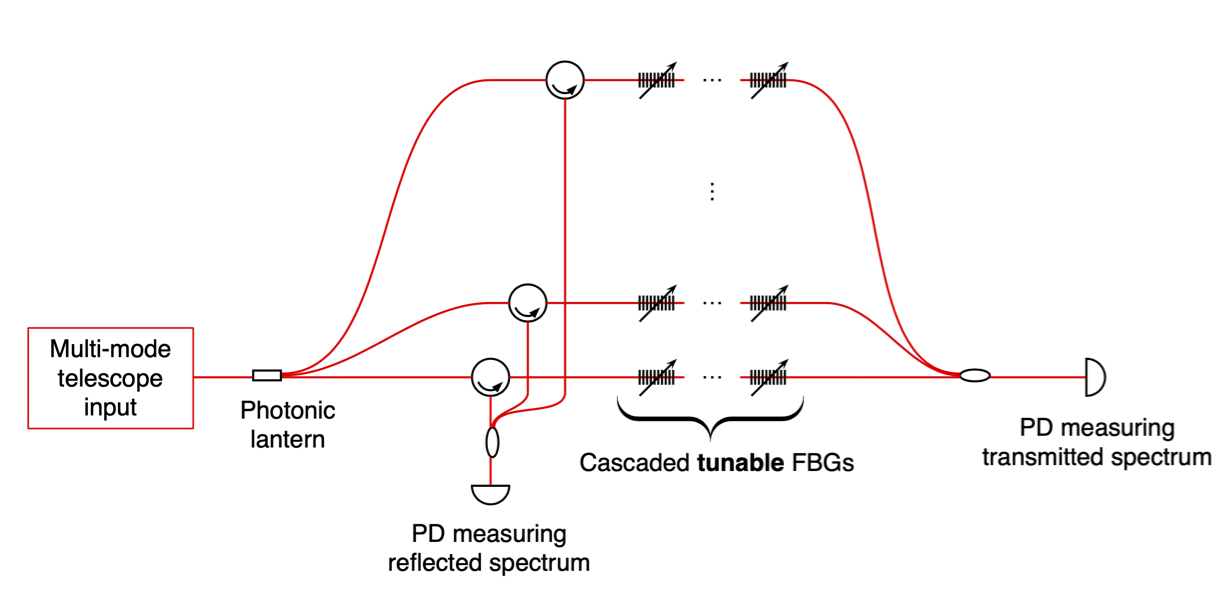}
   \end{tabular}
   \end{center}
   \caption[example] 
   { \label{fig:fbg} 
A schematic of the fibre version of the correlation sensor. The main elements are similar to the silicon photonic version where a photodiode (PD) measures either the transmitted and/or reflected spectrum. A circulator is used to direct the reflected spectrum to a photodiode. The primary difference is that the individual FBGs can be tailored to a specific absorption feature. The modulation is also done differently by straining the fibre with cascaded FBGs written using a femtosecond laser. In the diffraction-limited case, only a single fibre will be required. 
}
   \end{figure}

Fibre Bragg gratings (FBGs) have been a well-established photonic solution for OH skyglow removal in the near-infrared\cite{2015SPIE.9507E..0CE}. This involves the creation of custom FBGs that target specific airglow lines that suppress them to improve the sensitivity of low resolution near-infrared spectroscopy. Recent work has also shown similar Bragg gratings can be made in silicon photonics\cite{2021OExpr..2915867P}. The same concept can be applied to a correlation sensor but instead of holding the notch filters static, we are able modulate them across absorption features of our source by mechanically adjusting the strain of the optical fibre. The overall instrument concept is shown in Figure \ref{fig:fbg} where we cascade individual FBGs that are designed for a specific absorption feature. These FBGs are modulated simultaneously by stretching the fibre to obtain the correlation signal.  

Our implementation on this platform is less mature than the previous one, but we have been able to successfully write custom FBGs with shaped femtosecond laser pulses. The laser beam has been shaped into a blade to write periodic structures in fibre cores (Zavyalova et al., this conference) to construct FBGs. The fibres are strained using linear translation. A similar experimental setup as shown in Figure \ref{fig:siexp} was constructed to successfully detect CO$_2$ with the gas cell. The primary advantages of this method over the previously described platform are the ease of coupling, larger free spectral range, and notch filters that can be tailored to specific absorption features at high accuracy. The silicon photonic platform offers other advantages such as much more compact footprint, easier fabrication, and the ability integrate multiple photonic devices on a single chip. 

\section{Future Directions}
\label{sec:future}
\begin{figure} [ht]
   \begin{center}
   \begin{tabular}{cc} 
   \includegraphics[height=7cm]{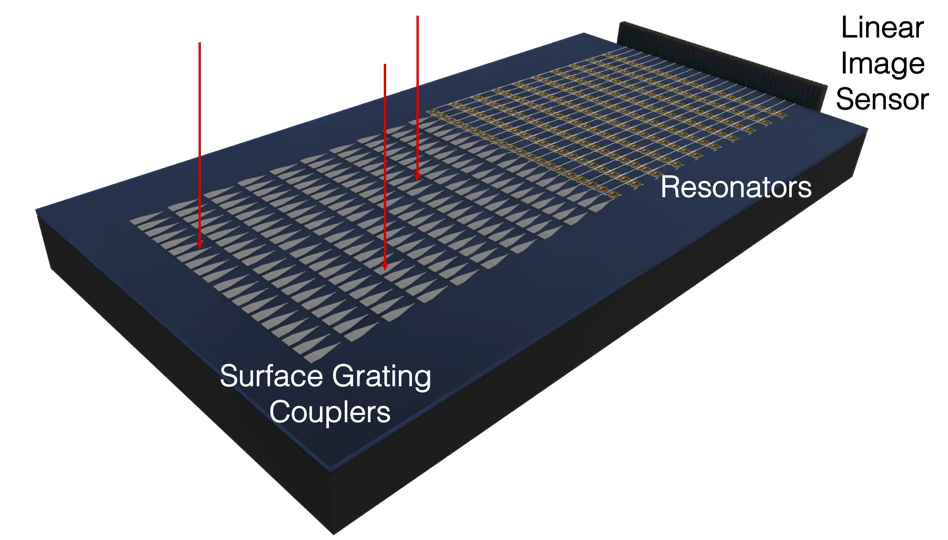}
   \end{tabular}
   \end{center}
   \caption[example] 
   { \label{fig:concept} 
A integrated concept for an exoplanet detection device behind an extreme AO system. The device will be located at the coronographic focal plane. The surface grating couplers will couple light from different spatial locations on the focal plane into the photonic device. The correlation sensor would be tailored to search for specific molecules across all spatial locations and its output will be captured by a linear photodiode array. The potential of this device is significant as it will have a much lower fabrication cost than a conventional integral field spectrograph while being easily replicable. }
   \end{figure}

While our immediate plans are to demonstrate the feasibility of this technique on key astronomical sources, namely solar system planets and bright stars, our overall goal is to make highly multiplexed and high dispersion spectroscopy much more affordable. There are two specific areas of significant potential in the next decade: 

One is the field of high contrast imaging spectroscopy for the detection of exoplanets. There is a significant push in the era of extremely large telescopes to combine high contrast imaging with high dispersion spectroscopy to obtain significant improvements in the star-planet contrast ratio. Tiling an extreme AO coronographic focal plane with an astrophotonic chip capable of detecting molecular signals from exoplanets (Figure \ref{fig:concept}) would be very powerful as it would not require a costly integral field, high dispersion spectrograph. 

The other is the field of multi-object spectroscopy. Since the age of GAIA, there is an ever increasing need to measure the RVs and compositions of stellar objects within our Galaxy. While there are optical and near-infrared surveys underway to do just that, our solution can over an alternate low cost solution, which could enable massively multiplexed multi-object spectroscopy. 

Lastly, one of the challenges that remains is the effective coupling of light from the telescope to the device. While traditional AO systems are able to solve this issue, we are also exploring low-cost photonic phase correctors, implemented using silicon photonics. These devices can correct for atmospheric turbulence and effectively couple the light into a single mode waveguide (Diab et al., this conference). This could effectively mitigate the coupling problem without requiring a complex AO system implemented at the observatory. 

\section{Summary}
\label{sec:summary}
We present a novel concept to carry out astrophysics directly on a photonic device by performing non-dispersive cross-correlation spectroscopy with a template transmission spectrum. By tailoring the transmission spectrum accordingly, we can measure RVs, compositions, as well as detect faint objects with known spectral signatures. While designed to solve specific scientific problems, the cost of manufacturing devices is very low, enabling these devices to be easily replicated and made for a host of science cases. Our initial devices have been able to successfully detect molecular gases in a lab environment, and in the coming year, we have an immediate goal to test them on astronomical sources. 

\acknowledgments 
 
S.S. acknowledges that the NSERC Discovery Grant, the Canada Foundation for Innovation, the Ontario Research Fund, and the Dunlap Institute support his research.  

\bibliography{report} 
\bibliographystyle{spiebib} 

\end{document}